\documentstyle[epsf,12pt,psfig]{article}
\pssilent

\begin{document}

\begin{titlepage}

\begin{centering}
\vspace*{.5in}
{\Large {\bf Linear Stability of Dilatonic Black
Holes$^*$}}
 
\vspace{1in}
{\bf Panagiota Kanti}
\\
\vspace{.4in}
{\it Division of Theoretical Physics,
Physics Department,}\\[2mm] {\it University of Ioannina,
Ioannina GR 451 10, Greece.} \\

\vspace{1in}
{\bf Abstract} \\
\vspace{.1in}
\end{centering}
{In this talk, we recall the most important features of the Dilatonic Black
Holes which arise in the framework of the Einstein-Dilaton-Gauss-Bonnet
theory and which are dressed with a classical long range dilaton field 
in contradiction with the existing ``no-hair" theorems of the Theory of
General Relativity. We demonstrate linear stability of these black hole 
solutions under small spacetime-dependent perturbations by making use of a 
semi-analytic method based on the Fubini-Sturm's theorem. As a result, the 
Dilatonic Black Holes constitute one of the very few examples of stable black 
hole solutions with non-trivial ``hair" that arise in the framework of a more 
generalised theory of gravity.}

\vspace{1.6in}

$^{*}$ {\footnotesize Talk presented at the ``International Workshop on Recent 
Developments in High Energy Physics" organized by the Hellenic Society
for the Study of High Energy Physics at Democritos NRC, Athens, Greece,
April 9-11, 1998.}   

\end{titlepage}

\section{Introduction} 
\paragraph{}
The Dilatonic Black Holes arise in the framework of the one-loop
corrected four-dimensional effective theory of the heterotic superstrings
at low energies. In ref.~\cite{kanti}, we have demonstrated numerically 
the existence of the Dilatonic Black Hole solutions with a regular event
horizon and an asymptotically flat behaviour at infinity in the presence
of higher-derivative gravitational terms such as the Gauss Bonnet (GB)
curvature-squared term. According to our analytic arguments~\cite{kanti},
it was the presence of this term which led to the by-passing of the
``no-scalar-hair" theorems~\cite{beken} and the existence of black hole 
solutions dressed with non-trivial classical dilaton hair.

The existence itself of the Dilatonic Black Holes is beyond any doubt
since our results were rederived and confirmed by other research groups
in subsequent works~\cite{pomaz}~\cite{maeda}. But, an important question
related to the nature and fate of our black hole solutions arises next~:
the question of stability. Although the ``no-hair" theorems refer to the
existence of black hole solutions and not to their stability, there is a 
tendency in the literature to reject some of these solutions due to their
instability under spacetime perturbations. At the same time, the conception
that the ``no-hair" theorems are indeed valid once we have assured the
stability of the black hole solutions has started to form among the
scientists. For these reasons, we need also to check the stability of our 
solutions, that is the Dilatonic Black Hole solutions~\cite{kanti}.

In the next paragraph, we are going to recall the most important features
of the Dilatonic Black Holes while in paragraph 3 we shall present
our semi-analytic method for the study of our solutions under linear 
spacetime-dependent perturbations~\cite{kanti1}. The key feature of our method
is the reduction of the time-dependent equations of motion of the theory
to a single one-dimensional Schr\"{o}dinger type differential equation.
As we shall see, the absence of bound states in the spectrum of this 
equation corresponds to the absence of unstable modes or equivalent to
the stability of our classical black hole solutions 
of the time-independent equations of the theory~\cite{kanti}. This linear
stability, if it comes out to be true, will be an outstanding result since the
corresponding stable black hole solutions will be unique in the 
framework of a generalised theory of pure gravity. 

\section{Dilatonic Black Holes} 
\paragraph{}
The action functional of the Einstein-Dilaton-Gauss-Bonnet (EDGB)
theory, in the context of which the Dilatonic Black Holes
have arisen, has the following form~\cite{kanti}
%%%%%%%%%%%
\begin{equation}
S=\int d^4 x \, \sqrt{-g} \left( -\frac{R}{2} - \frac{1}{4}
\partial_{\mu} \phi \partial^{\mu} \phi + \frac{\alpha' e^\phi}{8 g^2}
{\cal R}^2_{GB} \right)
\label{1}
\end{equation}
%%%%%%%%%%%
where
%%%%%%%%%%%
\begin{equation}
{\cal R}^2_{GB}=R_{\mu\nu\rho\sigma}R^{\mu\nu\rho\sigma}-
4R_{\mu\nu}R^{\mu\nu}+R^2
\end{equation}
%%%%%%%%%%%%%%%
is the well-known curvature squared Gauss-Bonnet term and $\phi$ stands
for the dilaton field. For simplicity, we have ignored all the other
scalar fields of the effective superstring theory, the axions $a$ and $b$ 
and the modulus field $\sigma$, as well as the gauge fields. 

The dilaton field and the Einstein's equations which follow from the action (\ref{1}) can be written in a covariant form in the following way
%%%%%%%%%%%%
\begin{equation}
\frac{1}{\sqrt{-g}}\,\partial _\mu [\sqrt{-g}\,\partial ^\mu \phi ]
=-\frac{\alpha '}{4g^2} \,e^\phi R^2_{GB}
\label{eq1}
\end{equation}
%%%%%%%%%%%
\vspace*{3mm}
%%%%%%%%%%%
\begin{equation}
R_{\mu\nu} - \frac{1}{2}\,g_{\mu\nu}\,R  =
- \frac{1}{2}\,\partial _\mu \phi
\,\partial _\nu \phi + \frac{1}{4} g_{\mu\nu}\,(\partial _\rho \phi )^2  -
\alpha ' \,{\cal K}_{\mu\nu}
\label{eq2}
\end{equation}
%%%%%%%%%%%%%
where
%%%%%%%%%%%%
\begin{equation}
{\cal K}_{\mu\nu}=(g_{\mu\rho}\,g_{\nu\lambda}+g_{\mu\lambda}\,g_{\nu\rho})
\,\eta^{\kappa\lambda\alpha\beta} D _\gamma
\,[{\tilde R}^{\rho\gamma}_{\,\,\,\,\,\alpha\beta} \,\partial _\kappa f]
\end{equation}
%%%%%%%%%%%%%%%
and
%%%%%%%%%%%%%%
\begin{equation}
\eta ^{\mu\nu\rho\sigma} = \frac{\epsilon ^{\mu\nu\rho\sigma}}
{\sqrt{-g}} \quad , \quad f=\frac{e^\phi}{8g^2} \quad , \quad 
\tilde{R}^{\mu\nu}_{\,\,\,\,\kappa\lambda}=\eta ^{\mu\nu\rho\sigma}
R_{\rho\sigma\kappa\lambda}\,\,.
\end{equation}
%%%%%%%%%%%%%%
\paragraph{}
We assume that the spacetime background is spherically symmetric and it
is described by the following line element 
%%%%%%%%%%%
\begin{equation}
ds^2=e^{\Gamma(r)} dt^2-e^{\Lambda(r)} dr^2 -r^2(d\theta^2
+\sin ^2\theta \, d\varphi^2)\,\,.
\label{2}
\end{equation}
%%%%%%%%%%%

If the above ansatz is to describe a black hole solution, we must demand
the following behaviour for the components of the metric tensor and the scalar
filed near the horizon of the black hole
%%%%%%%%%%%%
\begin{eqnarray}
e^{-\Lambda(r)}&=& \lambda_1 (r-r_h)+\lambda_2 (r-r_h)^2 + ... \nonumber \\[4mm]
e^{\Gamma(r)}&=&\gamma_1 (r-r_h) +\gamma_2 (r-r_h)^2+... 
\label{rr_h}\\[4mm]
\phi(r)&=&\phi_h+\phi'_h (r-r_h)+\phi''_h (r-r_h)^2+...\nonumber
\end{eqnarray}
%%%%%%%%%%%%

Substituting the above expansions in the equations of motion, we find that
the derivative of the dilaton field at the horizon satisfies the relation
%%%%%%%%%%%%%
\begin{equation} 
\phi_h'=\frac{g^2}{\alpha'}
r_he^{-\phi_h}\left(-1 \pm \sqrt{1-\frac{6(\alpha ')^2
e^{2\phi_h}}{g^4r_h^4}}\right)
\label{phiprime}
\end{equation}
%%%%%%%%%%%%%%
which results in the following constraint for the coupling function of
the dilaton field with the Gauss-Bonnet term
%%%%%%%%%%%%%% 
\begin{equation}
\frac{\alpha' e^{\phi_h}}{g^2} < \frac{r_h^2}{\sqrt{6}}\,\,.
\label{condition}
\end{equation}
%%%%%%%%%%%%%%
If the coupling function does not satisfy the above constraint, the solutions
of the equations of motion of the theory can no longer be described by the
concept of the black hole.
%%%%%%%%%%%%%%
At the other end of the radial space, that is in the limit $r \rightarrow
\infty$, one assumes the following asymptotically flat behaviour: 
%%%%%%%%%%%%%
\begin{eqnarray}
e^{\Lambda(r)}&=&1+\frac{2M}{r}+\frac{16M^2-D^2}{4r^2}+
O \left( \frac{1}{r^3}\right)
 \nonumber \\[4mm]
e^{\Gamma(r)}&=&1-\frac{2M}{r}+O \left( \frac{1}{r^3} \right) 
\label{rinfty} \\[4mm]
\phi(r)&=&\phi_\infty+\frac{D}{r}+\frac{MD}{r^2}+
O \left( \frac{1}{r^3} \right)  \nonumber 
\end{eqnarray}
%%%%%%%%%%%%%%
The constants $M$ and $D$ that appear in the above expressions stand for
the ADM mass and the dilaton charge of the black hole, respectively. 

If we integrate numerically the equations of motion (\ref{eq1})-(\ref{eq2})
by making use of the ansatz (\ref{2}) and the expansions
(\ref{rr_h}) and (\ref{rinfty}), we are led to a continuous
one-parameter family of regular black hole solutions. The solution for the scalar field and the metric components is displayed in Figures 1 and 2.
The main feature of the solution, that we are going to make use of during
the stability analysis, is the monotonic, non-intersecting behaviour of the
dilaton field from $r_h$ to infinity. 

\vspace*{5mm} 
%%%%%%%%%%%%
\centerline{\hbox{
$\psfig{figure=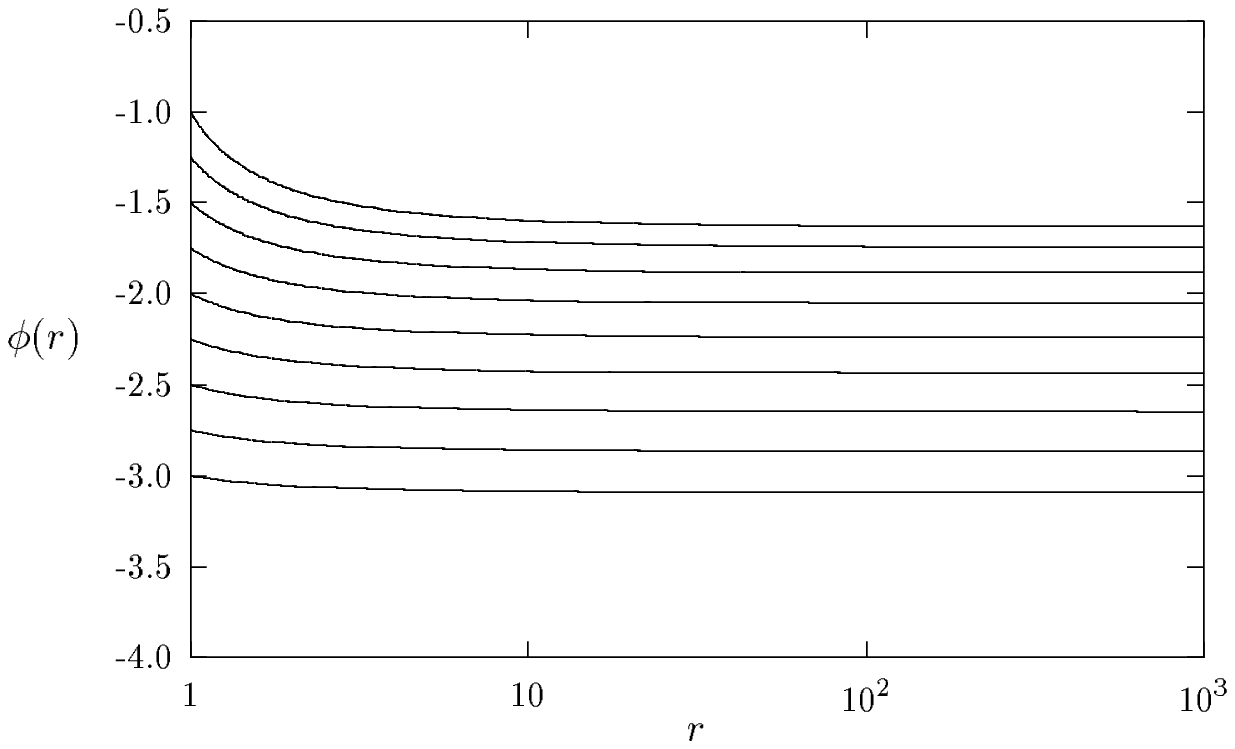,height=6cm,width=8cm}$
$\psfig{figure=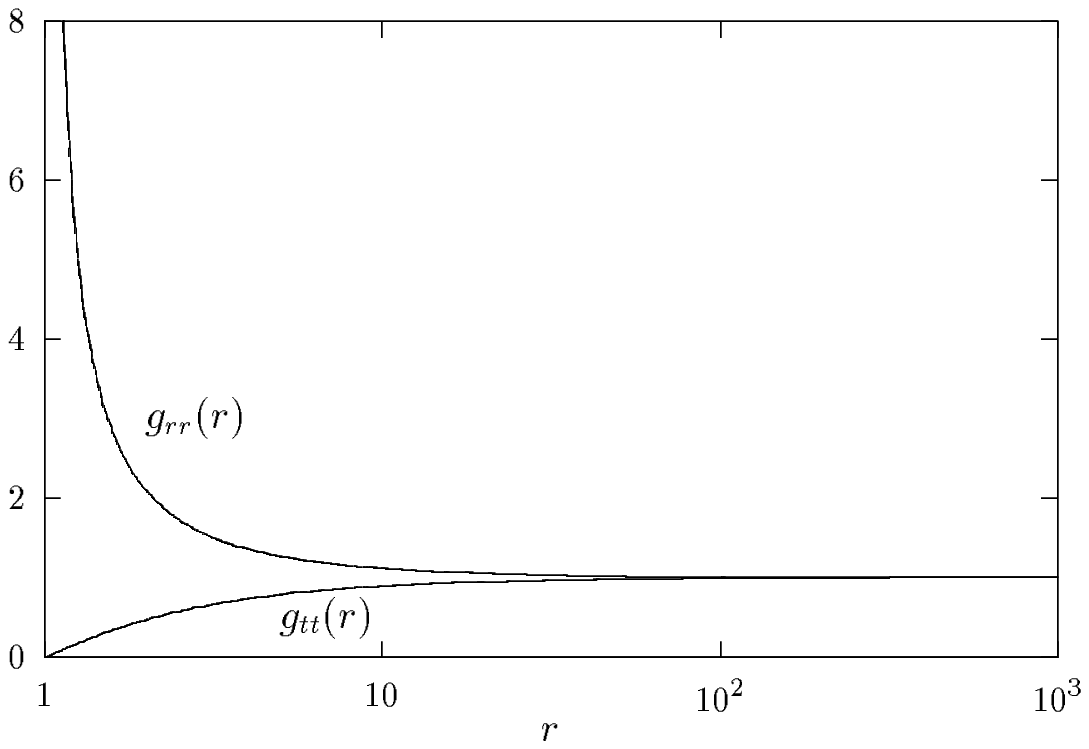,height=6cm,width=7cm}$}}
\begin{center}
{\small \hbox{\hspace*{-3cm}\vbox{\begin{tabular}{c}
{\bf Figure 1~:} Dilaton configurations for a \\ family of black hole
solutions, for\\ fixed $r_h=1$ and various values of $\phi_h$.
\end{tabular}} \hspace*{-8cm}
\vbox{\begin{tabular}{c} {\bf Figure 2~:} Metric components $g_{tt}$ and\\ 
$g_{rr}$ for the $\phi_h=-1$ and $r_h=1$\\ black hole solution.\\ \end{tabular}}}}
\end{center}
%%%%%%%%%%%%
\paragraph{}
The existence itself of these black hole solutions, of the Dilatonic Black
Holes, are in disagreement with the well-known ``no-hair" theorems of the
Theory of General Relativity. These theorems state that the only parameters
that may characterize a black hole are the mass $M$, the charge $Q$ and
the angular momentum $J$ and that the only long range fields that can
be associated with the black hole are the gravitational and the
electromagnetic ones. The Dilatonic Black Holes are clearly characterized
by another long range field, that of the dilaton field, a feature which is
prohibited by the Theory of General Relativity. As we mentioned before, it
is the presence of the Gauss-Bonnet term that causes the by-passing of the
``no-hair" theorems and the existence of the long range scalar field. On the
other hand, this disagreement with the ``no-hair" theorems is only partial
in the sense that we have not found any new conserved quantity apart from the
aforementioned ones. The dilaton charge $D$ has turned out to be a dependent on
the mass $M$ quantity, thereby leading to the secondary nature of the
dilaton ``hair"~\cite{wilczek}.
\vspace*{5mm}

\section{Linear Stability Analysis}
\paragraph{}
In this section, we are going to present our semi-analytic method for the
study of our solutions under linear spacetime-dependent
perturbations~\cite{kanti1}. For the needs of our analysis, we are going to 
assume that both of the metric components as well as the scalar field depend
not only on the radial coordinate $r$ but also on the time coordinate $t$. 
Moreover, according to the linear stability 
analysis~\cite{straumann}~\cite{linear}, we may write the three unknown 
functions of the problem as a sum of two parts in the following way 
%%%%%%%%%%%%
\begin{eqnarray}
 \Gamma (r,t) &=& \Gamma (r) + \delta \Gamma (r,t)=\Gamma (r) + 
\delta \Gamma (r) e^{i\sigma t} \nonumber \\[2mm]
 \Lambda (r,t) &=& \Lambda (r) + \delta \Lambda (r,t)=\Lambda (r) 
+ \delta \Lambda (r) e^{i\sigma t}
\label{perturbations} \\[2mm]
 \phi (r,t) &=& \phi (r) + \delta \phi (r,t)=\phi (r) + \delta \phi (r)
e^{i\sigma t} \nonumber
\end{eqnarray}
%%%%%%%%%%%%%%%%
The first part of each of the above functions is a purely radial part which
corresponds to the classical Dilatonic Black Hole solutions already found
in Ref.~\cite{kanti}. The second part is the product of a small radial part,
denoted by the prefactor $\delta$, by an harmonic function of time. These small
radial parts are the linear perturbations of our solutions. Note the appearance
of the parameter $\sigma$ in the above expressions. It is easy to understand
that for negative $\sigma^2$, that is for purely imaginary $\sigma$, even
small perturbations may be led to extremely large values at late times. For
this reason, we have to assure the absence of negative $\sigma^2$ in the case
of our black hole solutions.

If we substitute the expressions (\ref{perturbations}) in the time-dependent
equations of motion (\ref{eq1})-(\ref{eq2}) and make use of the harmonic
dependence of the perturbations on time $t$, we end up with a group of 
differential equations with respect to the radial coordinate $r$ for the
linear perturbations $\delta\Gamma$, $\delta\Lambda$ and $\delta\phi$. 
After some long and tedious algebraic computation, we manage to construct
a single differential equation for the dilaton perturbation $\delta\phi$ from
which the other two perturbations, $\delta\Gamma$ and $\delta\Lambda$,
have been eliminated. This differential equation has the following structure
%%%%%%%%%%%%%%%
\begin{equation}
A\,\delta\phi''+ 2B\,\delta\phi' + C\,\delta\phi +
\sigma^2\,E\,\delta\phi=0
\label{dphi1}
\end{equation}
%%%%%%%%%%%%%%%%%%
where $A$, $B$, $C$, and $E$ are rather complicated functions of
$\phi$, $\phi'$, $\phi''$, $\Lambda$, $\Lambda'$, $\Gamma$, $\Gamma'$
and $\Gamma''$. All of these coefficients are very well behaved near
infinity, that is in the limit $r \rightarrow \infty$, according to the 
following expressions
%%%%%%%%%%%%%%%%%
\begin{eqnarray}
A= 1+ O(\frac{1}{r^5})
\quad &,& \quad 
B=\frac{1}{r}+\frac{M}{r^2}+O(\frac{1}{r^4}) \label{infA} \\[5mm]
C= \frac{D^2}{2r^4}+O(\frac{1}{r^5}) \quad &,& \quad
E=1 + \frac{4M}{r}+\frac{4M^2}{r^2}+
O\left( \frac{1}{r^4}\right)\,\,.
\label{asinf1}
\end{eqnarray}
%%%%%%%%%%%%%%%%%
However, three of them, $B$, $C$ and $E$, take on infinite values near the 
horizon of the black hole, while $A$ remains finite, in the following way
%%%%%%%%%%%%%%%%%%%
\begin{eqnarray}
A & = & \frac{2\,\sqrt{x}}{1+\sqrt{x}}+ O\,(r-r_h) \quad , \quad
B  = \frac{\sqrt{x}}{1+\sqrt{x}}\,\frac{1}{(r-r_h)}+O\,(1)
\label{A}\\ [4mm]
C & = & \frac{2\,e^{2\phi_h}}{r_h^4\,(1+\sqrt{x}\,)}
\,\frac{1}{(r-r_h)^2}+O\left( \frac{1}{r-r_h} \right)
\label{C}\\[4mm]
E & = & \frac{r_h\,\sqrt{x}}{\gamma_1}\,\frac{1}{(r-r_h)^2}
+ O\left( \frac{1}{r-r_h} \right)\,\,.
\label{E}
\end{eqnarray}
%%%%%%%%%%%%%%%
where $x=1-6 e^{2\phi_h}/r_h^4$.

As a result, the differential equation (\ref{dphi1}) is not well defined at
one point only of the radial space, that is at the event horizon $r_h$. In
order to remove this singularity, it is necessary to introduce a new radial
coordinate, the so called ``tortoise" coordinate $r^*$, which is related to
the old one in the following way
%%%%%%%%%%%%
\begin{equation}
\frac{dr^{*}}{dr}=e^{-(\Gamma-\Lambda)/2}\,\,.
\label{tortoise}
\end{equation}
%%%%%%%%%%%%
The introduction of the new radial coordinate leads to the extension of the 
radial space $[\,r_h, \infty)$ over the entire real axis $(-\infty, \infty)$
in such a way that the relative slope of any two curves in Fig.~1 remains
unchanged. Nevertheless, the introduction of the new coordinate is powerful
enough to render all of the coefficients in the differential equation 
(\ref{dphi1}) finite over the entire radial space. Now, the perturbed
equation for the dilaton field takes the form
%%%%%%%%%%%%%%%
\begin{equation}
{\cal{A}}\,\frac{d^2\delta\phi}{dr^{*2}}+2{\cal{B}}
\frac{d\delta\phi}{dr^*}+({\cal{C}}+\sigma^2{\cal{E}})\,\delta\phi=0
\label{dphi2}
\end{equation}
%%%%%%%%%%%%%%
where
%%%%%%%%%%%%%
\begin{equation}
{\cal{A}}=A \quad,\quad {\cal{B}}=B\,e^{(\Gamma-\Lambda)/2}-
\frac{A}{4}\frac{d(\Gamma-\Lambda)}{dr^*} \quad , \quad
{\cal{C}}=e^{\Gamma-\Lambda} C \quad,\quad
{\cal{E}}=e^{\Gamma-\Lambda}E
\label{defABCE}
\end{equation}
%%%%%%%%%%%%

For reasons that will become clear later, we have to make another step in
order to eliminate the term in equation (\ref{dphi2}) which is proportional
to $\delta\phi'$. This step involves the use of the auxiliary function
$F$ defined as
%%%%%%%%%%%%%
\begin{equation}
F=\exp \left( \,\,\int _{-\infty}^{r^*}
\frac{{\cal{B}}}{{\cal{A}}}dr^{*'}\right)
\label{defF}
\end{equation}
%%%%%%%%%%%%%
and the definition of a new function $u$ as the product of the auxiliary
function by the dilaton perturbation, $u=F\,\delta\phi$. Then, the 
differential equation for $\delta\phi$ takes the form 
%%%%%%%%%%%%%%%
\begin{equation}
p_*^2 u+
\left[ \,\frac{{\cal C}}{\cal{A}}+\sigma^2 \frac{{\cal E}}{\cal{A}}
-\frac{{\cal{B}}^2}{{\cal{A}}^2} -
p_*\left(\frac{{\cal{B}}}{{\cal{A}}}\right)
\right] \,u=0
\label{dphi4}
\end{equation}
%%%%%%%%%%%%%%%
where
%%%%%%%%%%%%
\begin{equation}
p_*=\frac{d}{dr^*} \,\,.
\end{equation}
%%%%%%%%%%%%%

In this form, equation (\ref{dphi4}) is an ordinary Schr\"{o}dinger equation
with well defined coefficients over the whole radial space. At the same time,
this differential equation is an eigenvalue problem where different values of
the parameter $\sigma^2$ correspond to different eigenvalues. It is easy to
understand that, since the number of negative values of $\sigma^2$ is equal
to the number of the unstable modes of our black hole solutions, the absence 
of states with negative $\sigma^2$ in the above eigenvalue problem is
equivalent to the linear stability of the Dilatonic Black Holes. 

In order to demonstrate the absence of the aforementioned states from the
spectrum of the eigenvalue problem (\ref{dphi4}),
we are going to make use of the asymptotic behaviour of the
eigenfunctions $u$ at both limits of the radial space, $r \rightarrow
r_h$ and $r \rightarrow \infty$, and of the Fubini-Sturm's 
theorem~\cite{fubini}. We are going
to concentrate our attention on two different kinds of eigenfunctions, namely
on $u_0$ which corresponds to zero eigenvalue $\sigma^2=0$ and on $u_\sigma$
which corresponds to negative eigenvalues $\sigma^2<0$. We start by noting
that near the horizon of the black hole, that is in the limit $r \rightarrow
r_h$ or equivalently $r^* \rightarrow -\infty$, the differential equation 
(\ref{dphi4}) assumes the form
%%%%%%%%%%%
\begin{equation}
p_*^2u_{\sigma} + k^2u_{\sigma}=0
\end{equation}
%%%%%%%%%%%%
where
%%%%%%%%%%%%
\begin{equation}
k^2 \equiv 
\frac{2\gamma_1e^{2\phi_h}}{(1+\sqrt{1-6e^{2\phi_h}})
\sqrt{1-6e^{2\phi_h}}}
+ \sigma^2=k_0^2+\sigma^2
\label{eqhorison}
\end{equation}
%%%%%%%%%%%%%
and where we have made use of the definitions (\ref{defABCE}) and
the asymptotic expansions (\ref{rr_h}). It can be easily seen that, while
for $\sigma^2=0$ the corresponding eigenfunction $u_0$ takes on a constant,
non-zero value at the event horizon, for $-k_0^2<\sigma^2<0$ the only
acceptable value of $u_\sigma$,
in order to ensure the finiteness of $\delta\phi$, is~: $u_\sigma \sim 
e^{|k|r^*} \rightarrow 0$. Here, we have to add that the eigenfunctions
with $\sigma^2<-k_0^2<0$ are absent from the spectrum of the states since
they have been shown to violate the continuous, non-degenerate nature of
the unbound states~\cite{kanti1}. 

At the other end of the radial space, in the limit $r \rightarrow \infty$,
it can be easily seen from the expressions (\ref{infA})-(\ref{asinf1}) that
the differential equation (\ref{dphi4}) takes the simple form
%%%%%%%%%%%
\begin{equation}
p_*^2u_{\sigma} + \sigma^2 u_{\sigma}=0\,\,. 
\label{unstable} 
\end{equation}
%%%%%%%%%%%%%
In the same way, the only acceptable behaviour of the eigenfunction
$u$ for $\sigma^2<0$, in the limit $r \rightarrow \infty$, is~: $u_\sigma
\sim e^{-|\sigma|r^*}\rightarrow 0$ while for $\sigma^2=0$, the solution is 
$u_0=c_1\,r+c_2$. Note that only the eigenfunctions $u_\sigma$ that correspond
to negative values of the parameter $\sigma^2$ vanish
at both limits of the radial space. For this reason, only these eigenfunctions
correspond to {\it physical perturbations} of our black hole
solutions~\cite{straumann}. 
\paragraph{}
Now, we arrive at the final step of our stability analysis. This final step
involves the use of the Fubini-Sturm's theorem of ordinary differential
equations~\cite{fubini}. According to the theorem, we consider the following
two differential equations~:
%%%%%%%%%%%%
\begin{equation}
u_1'' +  (q_1-p_1^2-p'_1)\,u_1 = 0
\label{dif1}
\end{equation}
\begin{equation}
u_2'' + (q_2-p_2^2-p'_2)\,u_2 = 0
\label{dif2}
\end{equation}
%%%%%%%%%%%%%
If the coefficients of the above equations satisfy the following relation
%%%%%%%%%%%%
\begin{equation}
p_2' + p_2^2 - q_2 \le p_1' + p_1^2 - q_1
\label{nodecondition}
\end{equation}
%%%%%%%%%%%%%
throughout the interval $[a,b]$, 
then, between {\it any two consecutive zeroes of function $u_1$},
in the interval $[a, b]$, {\it there is at least one zero of
function $u_2$}.

If we apply the above theorem to the spectrum of bound states of a 
Schr\"{o}dinger type eigenvalue problem, we obtain the well-known
``node rule", according to which, if we arrange all the bound states in an
increasing order of eigenvalues, the $n$~-~th eigenfunction has $n-1$ nodes~\cite{messiah}. This means that the ground state in the spectrum
of the bound states, denoted by $u_b$, has no zeroes throughout the
interval $[a, b]$.
\paragraph{}
Finally, we consider the case where $u_1=u_b$ with $\sigma^2_b$ being the most
negative eigenvalue of the system and $u_2=u_0$ with $\sigma_0^2=0$. We
multiply equation (\ref{dif1}) by $u_2$ and equation (\ref{dif2}) by $u_1$,
subtract the two equations and integrate the result over the entire domain
of $r^*$. Then, we obtain~:
%%%%%%%%%%%%
\begin{equation}
(u_\sigma\,p_*u_0-u_0\,p_*u_b)\bigg|_{-\infty}^{\infty}=
(\sigma_0^2-\sigma_b^2)\int_{-\infty}^\infty
\frac{{\cal E}}{{\cal A}}\,u_b\,u_0\,dr^*
\label{final}
\end{equation}
%%%%%%%%%%%%

%%%%%%%%%%%%%
\begin{figure}
\begin{center}
$\psfig{figure=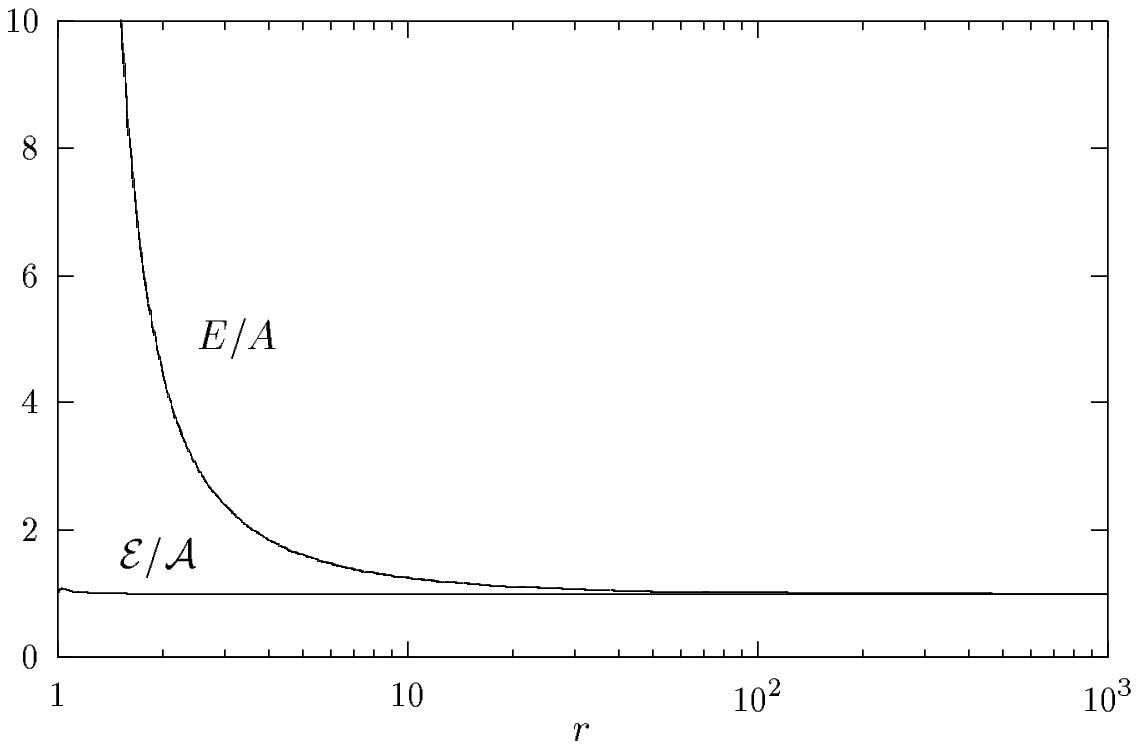,width=10cm,height=7cm}$\\[2mm]
{\bf Figure 3~:} The coefficient ${\cal E}/{\cal A}$ for the $r_h=1$ and
$\phi_h=-1$ black hole solution. The definite sign of this coefficient is
obvious. Also shown is the coefficient $E/A$ before the introduction of the
``tortoise" coordinate.
\end{center}
\end{figure} 
%%%%%%%%%%%%%

By making use of the asymptotic behaviour of the eigenfunctions $u_0$ and
$u_b$ at both limits of the radial space, a behaviour which was discussed
previously, it can be easily seen that the left hand side of the above equation
is zero. The same must also hold for the right hand side of this equation.
Note that the integral that appears on the right hand side contains the product 
of three functions, ${\cal E}/{\cal A}$, $u_b$ and $u_0$. The ratio of the
coefficients ${\cal E}$ and ${\cal A}$, after the introduction of the 
``tortoise" coordinate, is a well defined function over the entire radial
space as it is shown in Figure 3. Moreover, it is a positive definite 
function without any zeroes throughout the radial domain. On the other hand,
according to the ``node rule", the function $u_b$ is the ground state in the
spectrum of bound states of the eigenvalue problem and, as such, it has no
zeroes as well. The function $u_0$ demands some more attention  since it
corresponds to a vanishing value of the parameter $\sigma$. In this case,
the time dependence in the expressions (\ref{perturbations}) of the linear
perturbations disappears and the small radial parts can be absorbed into
the large radial parts which describe the classical Dilatonic Black Hole
solutions. As a result, the eigenfunction $u_0$ can be easily constructed
out of the difference of any two of the curves in Figure 1. The monotonic,
non-intersecting behaviour of these curves ensures the absence of zeroes
in the expression of $u_0$. Since each one of the three functions that
appear inside the integral has a definite sign, their product will have a
definite sign as well and the integral can never vanish. Then, the only
consistent case is the degenerate one $\sigma_b^2=\sigma_0^2=0$, which means
that, actually, the ground state of the spectrum is the one with the zero
eigenvalue and that all of the bound states with negative values of the
parameter $\sigma^2$ are absent. Since the number of negative values of
$\sigma^2$ is equal to the number of the unstable modes of our solutions,
the absence of bound states implies the linear stability of the Dilatonic
Black Holes.

\section{Conclusions}
\paragraph{}
By the use of a semi-analytic method, we have demonstrated the linear
stability of the Dilatonic Black Holes, which arise in the framework of
the Einstein-Dilaton-Gauss-Bonnet theory, under small spacetime-dependent
perturbations. As a result, our black hole solutions have no reason to be
rejected, even if we make use of the argument of stability, and the
concept of the validity of the ``no-hair" theorem among the stable solutions
has proven to be false. Our result is extremely important since it renders
our black hole solutions linearly stable and makes the long range dilaton
field, or equivalently the ``dilaton hair", one of the very few examples of stable, although of ``secondary type", hair that bypasses the ``no-hair"
theorems of the Theory of General Relativity.

The stability itself of the Dilatonic Black Holes under spacetime-dependent
perturbations is very important since the lifetime of these black hole
solutions becomes substantially longer. As a result, Dilatonic Black Holes,
that may have been formed in the past, are rather possible to have survived
until the present epoch. In this case, the possibility of detecting them is
not negligible at all and this might provide a test for the predictions of
superstring theory at low energies.

As a final remark, we would like to note that the Schwarzschild Black Hole,
which arises in the framework of the Theory of General Relativity, is also
linearly stable. This means that our black hole solutions and the Schwarzschild
Black Hole share the same kind of stability. Moreover, the Dilatonic Black
Holes arise in the framework of the Einstein-Dilaton-Gauss-Bonnet theory
which is a pure gravity theory--the dilaton field simply plays the role of the connecting link between the scalar curvature $R$ and the Gauss-Bonnet term.
Taking into account the above as well as the fact that these two black hole 
solutions are characterized by the same parameter at infinity, the mass $M$,
we are led to the conclusion that the Dilatonic Black Holes can be considered
as the most direct generalisation of the Schwarzschild Black Hole in the
framework of the effective superstring theory.

\end{document}